\documentclass[10pt]{article}
\usepackage[OE]{express}

\begin{document}
\title{Nonlinear refractive index of electric field aligned gold nanorods suspended in index matching oil measured with a Hartmann-Shack wavefront aberrometer}

\author{Melissa Maldonado,\authormark{1} Leonardo de S. Menezes,\authormark{1,*} Leonardo F. Araujo,\authormark{2} Greice K. B. da Costa,\authormark{3} Isabel C. S. Carvalho,\authormark{2} Jake Fontana,\authormark{4} Cid B. de Ara{\'u}jo,\authormark{1} and Anderson S. L. Gomes\authormark{1}}

\address{\authormark{1}Departamento de F{\'i}sica, Universidade Federal de Pernambuco, Recife-PE, 50670-901, Brazil\\
\authormark{2}Department of Physics, Pontif{\'i}cia Universidade Cat{\'o}lica do Rio de Janeiro (PUC-RIO), Rio de Janeiro, 22451-900, Brazil\\
\authormark{3}Departamento de F{\'i}sica, Universidade Federal Rural do Rio de Janeiro, Serop{\'e}dica-RJ, 23897-000, Brazil\\
\authormark{4}Naval Research Laboratory,~4555 Overlook Ave. Washington, D.C. 20375, USA}

\email{\authormark{*}lmenezes@df.ufpe.br}

\begin{abstract}

The capability to dynamically control the nonlinear refractive index of plasmonic suspensions may enable innovative nonlinear sensing and signaling nanotechnologies. Here, we experimentally determine the effective nonlinear refractive index for gold nanorods suspended in an index matching oil aligned using electric fields, demonstrating an approach to modulate the nonlinear optical properties of the suspension. The nonlinear optical experiments were carried out using a Hartmann-Shack wavefront aberrometer in a collimated beam configuration with a high repetition rate femtosecond laser. The suspensions were probed at $800~nm$, overlapping with the long-axis absorption peak of the nanorods. We find that the effective nonlinear refractive index of the gold nanorods suspension depends linearly on the orientational order parameter, $S$, which can be understood by a thermally induced nonlinear response. We also show the magnitude of the nonlinear response can be varied by $\sim$ 60\%.
\end{abstract}

\ocis{(190.4360) Nonlinear Optics, devices; (190.4870) Thermal effects; (350.6830) Thermal lensing; (120.3688) Lightwave analyzers; (250.5403) Plasmonics.}



\section{Introduction}

Metallic nanoparticles can confine and focus light below the diffraction limit, leading to interesting optical properties and applications \cite{Rocksthul2013}. The unique nonlinear optical properties of plasmonic nanoparticles hold significant promise for developing disruptive optical materials and devices. The nonlinear response of these materials may be well suited for chemical and biological sensing, optical limiters and switches, and nanoscale light sources \cite{Jiang2018,Chen2013,Ray2010}. Moreover, the potential to dynamically control the nonlinear response of these materials is a key element to opening up advanced application spaces.  Recently, external electric fields were demonstrated to dynamically control the orientational order and in turn the linear optical response of suspensions of gold nanorods at visible and near infrared wavelengths \cite{Fontana2016,Etcheverry2017,Etcheverry20172}. These color tunable suspensions enabled fast switching times and large signal modulations. However, the nonlinear response of such dynamic systems is not well characterized or understood. In ref. \cite{Padilha2009}, the Z-Scan technique \cite{SheikBahae1990} was used to demonstrate and briefly discuss the influence of gold nanorods alignment on the effective nonlinear refractive index, reporting {$n_2^{eff}$ values varying up to 40\% (ranging from 7.8 to 5.0 $\times 10^{-15}~cm^2/W$)} between the isotropic to partially aligned states, for light polarizations parallel and perpendicular to the applied electric field direction. 

Here experiments were carried out to understand the nonlinear optical response of electric field aligned gold nanorods using a Hartmann-Shack (or Shack-Hartmann) wavefront aberrometer in a collimated beam configuration and a high repetition rate femtosecond laser. We demonstrate the ability to tune the effective nonlinear refractive index of gold nanorods  suspended in an index matching oil by controlling their orientational order. We show by aligning the nanorods in an external electric field that magnitude of the effective nonlinear response can be varied by $\sim$ 60\% and is proportional to the orientational order parameter.

\section{Experimental results and discussion}

\subsection{Sample preparation and linear optical characterization}

To quantify the degree of electric field induced orientational order of the nanorods, an aqueous suspension of gold nanorods with length = $75~nm$ and diameter =$25~nm$ were acquired commercially (Nanopartz, A12-25-700-CTAB). The nanorods were coated with a polymer shell (Polymer Source, thiol-terminated polystyrene, $M_{n} = 50~k$), to enable the nanorods to be suspended in toluene and subsequently aligned using electric fields as described in ref. \cite{Fontana2016}. To eliminate solvent evaporation, preventing density changes, the nanorods were re-suspended in an index matching oil (Series A, Cargille index matching fluid, $n_{D} = 1.5700$). The index matching oil ($2 ~ml$) was placed into a $5~ml$ glass vial. Then $2~ml$ of the nanorod suspension ($2~nM$) was gently placed on top of the oil. The vial was placed in a fume hood for 2 days, allowing the toluene to completely evaporate, dispersing the nanorods into the oil.

Figure 1(a) is the experimental setup to verify the relationship between the absorption and degree of alignment of the gold nanorods suspended in the oil as a function of electric field, ensuring the oil does not alter the response. The nanorod-oil suspension was placed into a cuvette (C), composed of two $1~mm$ thick microscope slides glued together with a $1~mm$ glass spacer. On the outside of each microscope slide indium tin oxide pads were chemically etched and served as electrodes to apply the electric field $E$ across the suspension. To prevent dielectric breakdown of the air while applying the electric field, the cuvette was put in a tank (O) of transformer oil (Lubrax AV 24), with a thickness $T=34~mm$. The voltage $V$ was applied to the cuvette using a sinusoidal function generator (SRS DS345, 60 Hz) coupled to a $2000\times$ voltage amplifier (Trek 20/20C). The entire alignment system (AS) is composed of tank with transformer oil, cuvette with voltage pads and the gold nanorods suspension. Numerical simulations of the system were carried out using the COMSOL Multiphysics software to calculate the electric field inside the cuvette for a given applied voltage. From the simulations a relationship of $E =556V$ was retrieved ($E$ given in Volts/meter and $V$ given in Volts). In this computational study, we observed negligible fringe field effects, i.e., the electric field amplitude close to the glass-sample interface is only slightly higher than that at the center of the cuvette, with 0.005\% difference. An unpolarized white light (WL) source (Ocean Optics HL-2000-LL) was used to probe the extinction of the nanorod suspension in the geometry shown in Fig. 1(a), where the wavevector of the probe light is parallel to the applied electric field direction. The collected light was analyzed by using a spectrometer (SP, Ocean Optics HR4000).

\begin{figure}[htbp]
\centering\includegraphics[width=13cm]{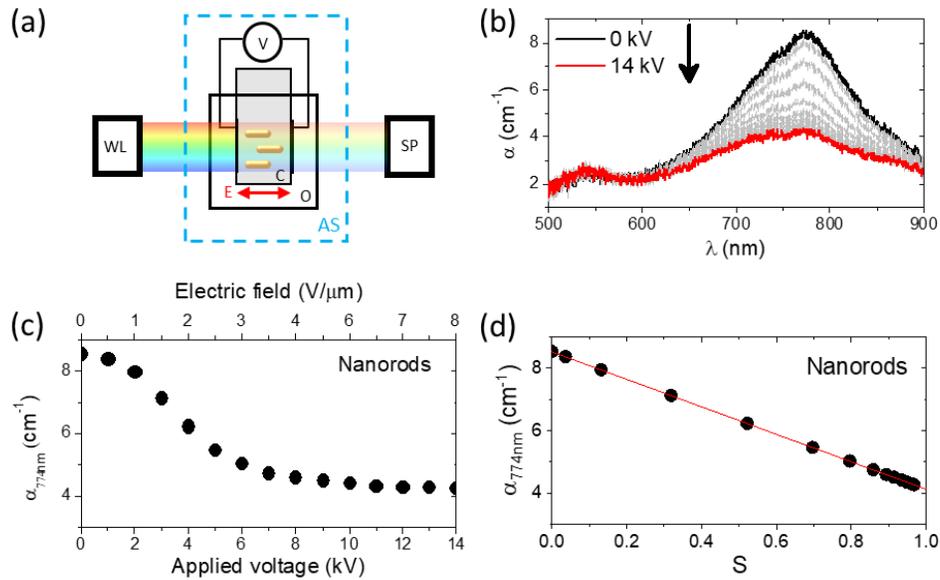}
\vspace{-1cm}
\caption{(a) Experimental gold nanorods alignment setup. (b) Extinction spectra of the gold nanorods suspension as a function of applied voltage. (c) Extinction at $774~nm$ (peak of the plasmon resonance associated with the LSP) versus applied voltage. (d) Extinction at the LSP resonance as a fucntion of orientational order parameter $S$. The solid red line is a linear fit, for which $A=8.5~cm^{-1}$, $B=-4.4~cm^{-1}$ and $E_{c}=1.2~V \cdot \mu m^{-1}$ (see Eqs. (1) and (2)). In Figs. (c) and (d), the error bars are smaller than the points representing the experimental data.}
\label{1}
\end{figure}

The evolution of the extinction spectrum of the gold nanorods was analyzed as a function of applied voltage as shown in Fig. 1(b). Initially, the gold nanorods are randomly oriented, giving rise to two characteristic absorption peaks, one from the longitudinal surface plasmon (LSP) at $774~nm$ and the other from the transverse surface plasmon (TSP) at $538~nm$. For the experimental geometry, as the voltage is increased ($0$ to $14~kV$), the LSP peak decreases by $\approx50\%$ since the nanords are being probed parallel to the aligning field direction, while the TSP peak slightly increased and is smaller in magnitude. 

The change in the magnitude of the LSP absorption peak as a function of applied voltage is shown in Fig. 1(c). The extinction coefficient, $\alpha$, is proportional to the orientational order parameter, $S$ \cite{Fontana2016}, \begin{equation}\alpha=A+BS
\end{equation}\
where $A$ and $B$ are constants and
\begin{equation}
S=\frac{\int _{0}^{1}\frac{1}{2}\left (3\cos ^{2}\left (\theta \right ) -1\right )e^{\genfrac{(}{)}{}{}{E}{E_{c}}^{2}\cos ^{2}\left (\theta \right )}d\left (\cos \left (\theta \right )\right )}{\int _{0}^{1}e^{\genfrac{(}{)}{}{}{E}{E_{c}}^{2}\cos ^{2}\left (\theta \right )}d\left (\cos \left (\theta \right )\right )}
\end{equation}
where $\theta$ is the angle between the gold nanorods long axis and the applied electric field direction, ${E}$, and $E_{c}$ is the critical electric field needed to align the nanorods. If $S=0$, then the nanorods in the suspension are randomly oriented, and if $S=1$, then all nanorods are aligned in the applied electric field direction. The experimentally measured $\alpha$ at the LSP peak wavelength as a function of $E$ is fit to Eq. (1), determining the parameters $A$, $B$ and $E_{c}$, using a least square fitting algorithm. Figure 1(d) shows the extinction at the longitudinal peak wavelength versus the orientational order parameter $S$, demonstrating that the extinction due to the nanorods for these suspensions is proportional to the orientational order parameter, in agreement with ref. \cite{Fontana2016} and verifying that the index matching oil does not alter the electric field induced alignment of the gold nanorods.

\subsection{Hartmann-Shack wavefront aberrometer and nonlinear optical measurements}

While the linear optical response of the gold nanorods as function of orientational order parameter is well understood, to the best of our knowledge, there are no detailed studies of the nonlinear response versus nanorod alignment.

The effective nonlinear refractive index of the gold nanorods, $n_2^{eff}$, was measured using the experimental setup depicted in Figure 2(a). The same alignment system ``AS" in Fig. 1(a) was incorporated into Fig. 2(a) to align the gold nanorods. The optical nonlinearity due to the gold nanorods, as a function of their alignment, was probed with a Ti:Sapphire laser (Coherent Mira, $800~nm$, $150~fs$, $76~MHz$). The laser light first passes through a halfwave plate ($\lambda/2$), then through a Glan-Laser polarizer (GL) to control the laser beam incident energy. In order to probe the nanorod suspension in the cuvette, a telescope (T1) was used to reduce the diameter of the laser beam to $1.1~mm$. A second telescope (T2) was used to expand the laser light for illumination of the entire CCD detector of a Hartmann-Shack wavefront aberrometer (HSWA, Thorlabs WFS150C).

The HSWA is a CCD consisting of groups of $i\times j$ arrays (bins). In front of the CCD there is a microlens array, with one microlens per bin. For an incident plane wave, each microlens focuses the incoming light into the center of the respective bin. For an incident beam with arbitrarily shaped wavefront, a computer code checks the position of the microlens focus for each bin, as well as the light intensity into each bin. The code then reconstructs the beam wavefront based on the position and intensity of light in each bin using a set of orthogonal Zernike polynominal functions, which serve as basis for decomposing any wavefront \cite{Lakshminarayanana2011}. The second radial order Zernike polynominal, $Z_{2}^{0}$, is related to the wavefront focusing/defocusing and thus can be used to determine the medium nonlinear optical properties, \textit{i.e.}, those depending on the intensity of the laser light. Therefore determining $Z_{2}^{0}$ alone is enough information to characterize the wavefront distortions caused by nonlinear effects. Moreover, it has been shown in \cite{Rativa2009} that the other Zernike polynomials are not influenced significantly when inducing nonlinear curvature to the beam wavefront.

\begin{figure}[htbp]
\centering\includegraphics[width=13cm]{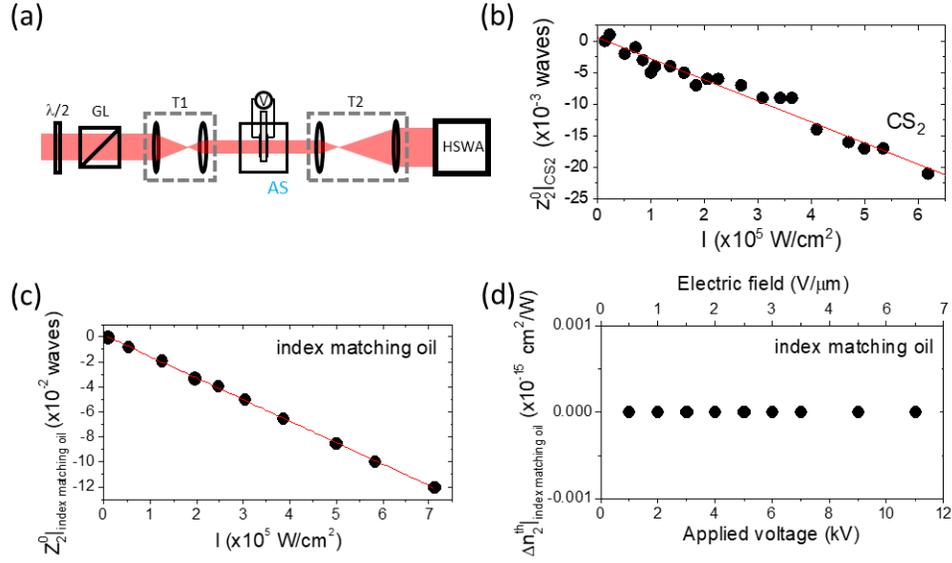}
\vspace{-1cm}
\caption{(a) Experimental setup to determine the effective nonlinear refractive index of the nanorods with a Hartmann-Shack wavefront aberrometer in a collimated beam configuration. (b) $Z_{2}^{0}\vert _{CS_{2}}$ versus laser peak intensity for liquid carbon disulfide ($CS_2$). (c) $Z_{2}^{0}\vert _{index~matching~oil}$ versus laser peak intensity for the index matching oil. (d) $\Delta n_{2}^{eff}\vert _{index~matching~oil} =n_{2}^{eff}\left (V\right ) -n_{2}^{eff}\left (V =0\right )$ versus applied voltage for the index matching oil. In Figs. (b), (c) and (d), the error bars are smaller than the points representing the experimental data.}
\label{2}
\end{figure}

The effective nonlinear refractive index $n_{2}^{eff}$ can be obtained from the expression \cite{Rativa2009}
\begin{equation}Z_{2}^{0} =\frac{2\pi L_{eff}n_{2}^{eff}}{f\lambda }I
\end{equation}
where $I$ is the laser peak intensity right before the sample, $L_{eff} =\left [1 -\exp \left (-\alpha L\right )\right ]/\alpha$ is the effective path length, $\alpha$ is the extinction coefficient at the excitation laser wavelength, $\lambda$ is the wavelength of light and $f$ is an adimensional factor related to the effective excitation intensity seen by the gold nanorods suspension. From Eq. (3), by measuring $Z_{2}^{0}$ versus $I$ and the resulting slope, the effective nonlinear refractive index, $n_{2}^{eff}$, can be determined.

Liquid carbon disulfide ($CS_{2}$), which has a well characterized $n_{2}=-4.4\times 10^{-15}~ cm^2/W$ at $800~nm$ and at the present excitation conditions\cite{Rativa2009}, was used to determine $f$ in Eq. (3).  In \cite{Gnoli2005}, the authors showed that the origin of the nonlinear behavior of $CS_2$ went from electronic at low repetition rate excitation pulses to thermal at high ones. Additionally, using a high repetition rate pumping scheme, in \cite{Singhal2017} the authors demonstrated that by flowing the nonlinear liquid the nonlinearity response is electronic in nature, whereas by reducing the flow rate, thermal nonlinearity starts playing the dominant role. In the present case, using a $2~mm$ thick quartz cuvette containing $CS_2$, $Z_{2}^{0}\vert _{CS_{2}}$ was determined as a function of the laser intensity (by controlling the pulse energy). From the resulting slope of the data in Fig. 2(b), $f$ was determined using Eq. (3) to be $1.9\times10^{-3}$. We note that the theoretical factor relating average and peak powers (or intensities) in the present experiments is $f_{theory}=1.7\times10^{-3}$. Considering that in Eq. (3) $Z_2^0\propto(I/f)$ and $I$ is the peak intensity, one gets an indication that the gold nanorods nonlinearity sees the pulsed excitation as a CW one due to the effective nonlinearity physical origin, such that the use of average powers or intensities may also be used for describing the system behavior, by using $f=1$ in Eq. (3). The reason for this will become clear when discussing the physical origin of the gold nanorods effective optical nonlinearity, below.

The shape of the wavefront measured by the HSWA is given by 
\begin{equation}W_{Total}=W_0+W_{L}+W_{NL}
\end{equation}
where $W_0$ is the background contribution from the experimental setup, $W_{L}$ is the linear contribution, and $W_{NL}$ is the nonlinear contribution. In terms of what the HSWA measures for the wavefront focus/defocusing, one writes 
\begin{equation}Z_2^0\vert _{Total}=Z_2^0\vert _{0}+Z_2^0\vert _{L}+Z_2^0\vert _{NL}
\end{equation}

To isolate the response of the gold nanorods from the transformer and index matching oils, a $34~mm$ thick tank filled with transformer oil and a $1~mm$ thick cuvette also containing transformer oil were inserted and probed with the  HWSA setup in Fig. 2(a). For each laser intensity, the respective $Z_2^0\vert_{Total}$ was determined. Then the transformer oil inside the cuvette was replaced with index matching oil. The measurement was repeated and a new $Z_2^{0~'}\vert_{Total}$ was obtained for each laser intensity. The contribution from the index matching oil can then be determined, 
\begin{equation}Z_2^0\vert_{Total~index~matching~oil}=Z_2^{0~'}\vert_{Total}-Z_2^0\vert_{Total}
\end{equation}
and is shown in Fig. 2(c). By fitting Eq.(3) to the data in Fig. 2(c), $n_{2}^{eff}\vert _{index~matching~oil}$ was determined to be $-8.8\times 10^{-15}~cm^2/W$.

To ensure that effective nonlinear refractive index of the oil does not change with the applied voltage, $n_{2}^{eff}\vert _{index~matching~oil}$ was measured as a function of applied voltage.  At a fixed laser intensity ($I=4.5 \times 10^{5}~W/cm^2$), a voltage was applied across the cuvette contaning the index matching oil ranging from $V =0-12~kV$, yielding a $n_{2}^{eff}\vert_{Total~index~matching~oil} \left (V\right )$ at each voltage. The change in the effective nonlinear refractive index as a function of voltage, 

\begin{equation}\Delta n_{2}^{eff}\vert _{index~matching~oil} =n_{2}^{eff}\vert _{index~matching~oil}\left (V\right ) -n_{2}^{eff}\vert _{index~matching~oil}\left (V =0\right )
\end{equation}
is reported in Fig. 2(d).

With the response of the host media characterized, the gold nanorods in the index matching oil were placed into the cuvette and measured without a voltage applied, $n_{2}^{eff}\vert _{gold~nanorods}\left (V =0\right )$. $Z_2^0\vert_{Total~gold~nanorods}$ was measured at $I=4.5\times 10^5~W/cm^2$ and subtracted from $Z_2^0\vert_{Total~index~matching~oil}$, to give $n_{2}^{eff}\vert _{gold~nanorods}\left (V =0\right )= -1.0\times 10^{-13}~cm^2/W$.

The orientationally dependent effective nonlinear refractive index of the gold nanorods suspension, 
\begin{equation}\Delta n_{2}^{eff}\vert _{gold~nanorods}=n_{2}^{eff}\vert _{gold~nanorods}\left (V \right )-n_{2}^{eff}\vert _{gold~nanorods}\left (V =0\right )
\end{equation}
was finally determined as a function of applied voltage and is reported in Fig. 3(a).

The relationship between the effective nonlinear refractive index and the orientational order of the gold nanorods is not known. However, if we assume that the nonlinear response of the gold nanorods is predominately thermally induced (Joule heating) due to the high repetition rate of the laser and that the slow thermal response is related to the orientationally dependent absorption cross section of the nanorods, then the nonlinear refractive index of the gold nanorods can be described as
\begin{equation}\Delta n_{2}^{eff}\vert_{gold~nanorods}=\Bigg(\frac{n_{2}^{eff,\parallel}+2n_{2}^{eff,\perp}}{3}\Bigg)-\frac{1}{3}\big(n_{2}^{eff,\parallel}-n_{2}^{eff,\perp}\big)S
\end{equation}
where $n_{2}^{eff,\parallel}$, $n_{2}^{eff,\perp}$ are the parallel and perpendicular effective nonlinear refractive indices for the longitudinal and transverse axis of the nanorod, respectively. If $S=0$, then the nanorods in the suspension are randomly oriented and $\Delta n_{2}^{eff}\vert_{gold~nanorods}=\frac{1}{3}\big(n_{2}^{eff,\parallel}+2n_{2}^{eff,\perp}\big)$, the isotropic value. Here, if $S=1$, then all the nanorods are aligned in the applied electric field direction, and for the specific experimental geometry, one obtains $\Delta n_{2}^{eff}\vert_{GNRs}=n_{2}^{eff,\perp}$. Therefore, the effective nonlinear refractive index of the gold nanorods is posited to depend linearly on the orientational order parameter,

\begin{equation}\Delta n_{2}^{eff}\vert_{gold~nanorods} =A' + B' S
\end{equation}
where $A'=\frac{1}{3}\big(n_{2}^{eff,\parallel}+2n_{2}^{eff,\perp}\big)$ and $B'=-\frac{1}{3}\big(n_{2}^{eff,\parallel}-n_{2}^{eff,\perp}\big)$.

To validate this relationship in Eq. (10), the effective nonlinear refractive index of the gold nanorods as function of applied electric field was fit using the same algorithm developed from Eq. (1), Fig. 3(a). In Fig. 3(b), one finds that the effective nonlinear refractive index of the nanorods depends linearly on the orientational order parameter, confirming the relationship in Eq. (10), where  $A' =0.0~cm^{2}/W$ and $B' =8.6 \times 10^{-16}~cm^{2}/W$.  The critical electric field retrieved from fitting is $E_{c} =1.2~V \cdot \mu m^{-1}$, in agreement with the value retrieved from the linear experiments, Fig. 1. As the amplitude of the applied electric field increases, in Fig. 3(b), the gold nanorods align along the laser propagation direction, leading to a decreasing LSP cross section. As a result, the induced current in the gold nanorods decreases and the thermal effects due to the Joule effect are reduced, making the gold nanorods effective nonlinear refractive index less negative and the sample less nonlinear.

\begin{figure}[htbp]
\centering\includegraphics[width=13cm]{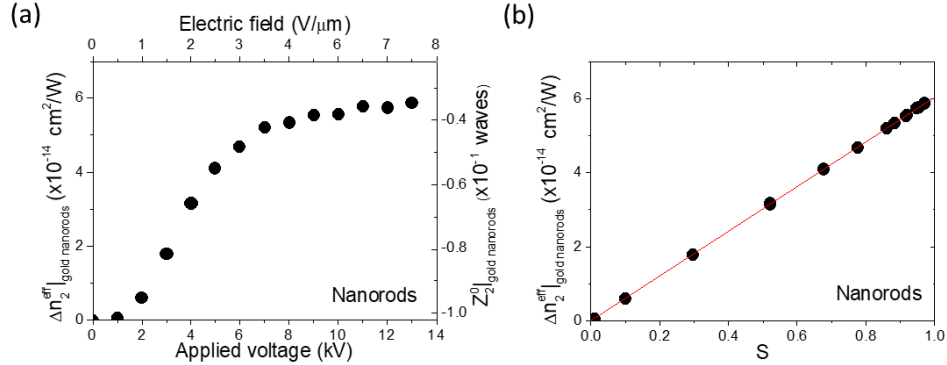}
\caption{(a) The Zernike polynominal, $Z_{2}^{0}\vert _{Total~gold~nanorods}$ for the gold nanorods in the dispersion as a function of applied voltage at a fixed laser intensity $I =4.5 \times 10^{5}~W/cm^{2}$ at $800~nm$, right hand side. The gold nanorods effective nonlinear refractive index (deviation from the case with no applied voltage) was calculated from $Z_{2}^{0}\vert _{Total~gold~nanorods}$ using Eq. (3), left hand side. (b) The gold nanorods effective nonlinear refractive index (deviation from the case with no applied voltage) at $800~nm$ as a function of the orientational order parameter The solid red line is a linear fit. In the figures, the error bars are smaller than the points representing the experimental data.}
\label{3}
\end{figure}

\section{Conclusion}

Experiments were carried out to characterize the nonlinear optical response of electric field aligned plasmonic nanorods using a Hartmann-Shack wavefront aberrometer in a collimated beam configuration and a high repetition rate femtosecond laser ($76~MHz$, $150~fs$) tuned to $800~nm$. We demonstrated the ability to dynamically control the effective nonlinear refractive index of a gold nanorods suspension in a index matching oil. We showed by aligning the nanorods in an external electric field that the effective nonlinear refractive index of the nanorods is proportional to the orientational order parameter, $S$, and  the magnitude of the nonlinear response can be varied by $\sim$ 60\%.  These results provide a straightforward means to predict and understand the nonlinear response of these materials as a function of orientational order, which may lead to novel nonlinear nanotechnology-based applications.

\section*{Funding}

This work was supported by the National Institute of Photonics (INFo) project; The support from the Conselho Nacional de Desenvolvimento Cient{\'\i}fico e Tecnol{\'o}gico-CNPq and Funda{\c c}{\~a}o de Amparo {\`a} Ci{\^e}ncia e Tecnologia do Estado de Pernambuco-FACEPE is acknowledged; I.C. and A.G. acknowledge support by the Office of Naval Research Global under ONRG-NICOP-N62909-16-1-2180; J. F. thanks the Office of Naval Research for support.

\end{document}